\pdfoutput=1
\documentclass[11pt]{article}
\usepackage[final]{acl}
\usepackage{times}
\usepackage{latexsym}
\usepackage[T1]{fontenc}
\usepackage[utf8]{inputenc}
\usepackage{microtype}

\usepackage{inconsolata}

\usepackage{graphicx}

\usepackage{booktabs}
\usepackage{array}

\usepackage{graphics,color,psfrag}
\usepackage{caption}   
\usepackage{bm}
\usepackage{url}
\usepackage{booktabs}

\usepackage[english]{babel}
\usepackage[linesnumbered,algoruled,boxed,lined]{algorithm2e}
\usepackage{multirow}
\usepackage{enumerate}
\usepackage{graphicx}
\usepackage{latexsym}
\usepackage{amsmath}
\usepackage{amssymb}
\usepackage{amsthm}
\usepackage{makecell}
\usepackage{lipsum}
\usepackage{cases}
\usepackage{epic}
\usepackage{color, colortbl}

\title{KG-RAG: Enhancing GUI Agent Decision-Making via Knowledge Graph-Driven Retrieval-Augmented Generation}

\author{
    Ziyi Guan$^{1,2}$\thanks{These authors contributed equally.}\thanks{Internship with Huawei Hong Kong Research Center.},
    Jason Chun Lok Li$^{1}$\footnotemark[1],
    Zhijian Hou$^{1}$\footnotemark[1],
    Pingping Zhang$^{3}$,
    Donglai Xu$^{5}$, \\
    \bfseries Yuzhi Zhao$^{1}$\thanks{Corresponding Author and Project Lead.},
    \bfseries Mengyang Wu$^{1}$,
    \bfseries Jinpeng Chen$^{3}$,
    \bfseries Nguyen Thanh Toan$^{1}$,
    \bfseries Pengfei Xian$^{1}$,\\
    \bfseries Wenao Ma$^{1}$,
    \bfseries Shengchao Qin$^{3,4}$,
    \bfseries Graziano Chesi$^{2}$,
    \bfseries Ngai Wong$^{2}$
    \\[2pt]
    $^{1}$Huawei Hong Kong Research Center 
    $^{2}$The University of Hong Kong \\
    $^{3}$City University of Hong Kong
    $^{3}$Guangzhou Institute of Technology, Xidian University \\
    $^{4}$ICTT and ISN Laboratory, Xidian University 
    $^{5}$Independent Researcher \\
    \texttt{zyguan@eee.hku.hk; yzzhao2-c@my.cityu.edu.hk}
}

\begin{document}
\maketitle

\begin{abstract}
Despite recent progress, Graphic User Interface (GUI) agents powered by Large Language Models (LLMs) struggle with complex mobile tasks due to limited app-specific knowledge. While UI Transition Graphs (UTGs) offer structured navigation representations, they are underutilized due to poor extraction and inefficient integration. We introduce KG-RAG, a Knowledge Graph-driven Retrieval-Augmented Generation framework that transforms fragmented UTGs into structured vector databases for efficient real-time retrieval. By leveraging an intent-guided LLM search method, KG-RAG generates actionable navigation paths, enhancing agent decision-making. 
Experiments across diverse mobile apps show that KG-RAG outperforms existing methods, achieving a 75.8\% success rate (8.9\% improvement over AutoDroid), 84.6\% decision accuracy (8.1\% improvement), and reducing average task steps from 4.5 to 4.1. Additionally, we present KG-Android-Bench and KG-Harmony-Bench, two benchmarks tailored to the Chinese mobile ecosystem for future research. Finally, KG-RAG transfers to web/desktop (+40\% SR on Weibo-web; +20\% on QQ Music-desktop), and a UTG cost ablation shows accuracy saturates at $\sim$4h per complex app, enabling practical deployment trade-offs.
\end{abstract}

\section{Introduction}
\begin{figure*}[t!]
    \centering
    \includegraphics[width=1\linewidth]{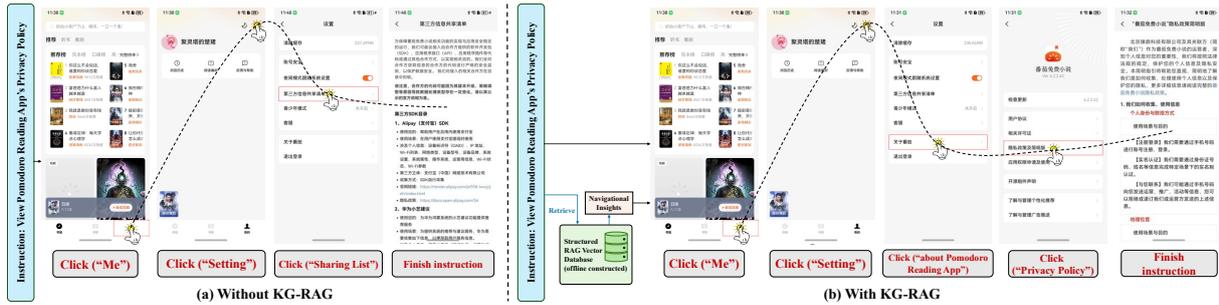}
    \caption{Improved Task Execution for ``View Privacy Policy'' in Tomato Novel App Using Graph-based RAG. \textbf{(a) Without KG-RAG:} Fails to identify the correct navigation path to access the ``View Privacy Policy'' page. \textbf{(b) With KG-RAG:} Successfully generates the correct path by leveraging knowledge graph-based information. }
    \label{fig:motivation}
\end{figure*}

In recent years, LLM-based GUI agents~\cite{zhang2023appagent, lee2023explore, yoon2023autonomous, hong2024cogagent, you2024ferret, wen2024autodroid, wang2025mobile, qin2025ui} have advanced in interacting with and navigating mobile applications. However, efficiently completing complex tasks remains challenging due to their limited app-specific knowledge, particularly when faced with unfamiliar user interfaces (UIs) and unconventional navigation logic. For example, as shown in Figure~\ref{fig:motivation}, the ``Privacy Policy'' page is deeply embedded within the ``About Pomodoro'' page, making it difficult to locate without a clear user menu, even for human users encountering the app for the first time. App UI Transition Graphs (UTGs), structured representations of app navigational flows, have emerged as promising solutions to enhance agents' navigational capabilities. Despite their potential, UTGs face significant barriers, including low-quality graph extraction from apps and inefficient integration into real-time decision-making processes.

To address the limitations of existing GUI agents, we propose \textbf{KG-RAG}, a \textbf{K}nowledge \textbf{G}raph-driven \textbf{R}etrieval-\textbf{A}ugmented \textbf{G}eneration framework designed to transform incomplete UTGs into structured vector databases, enabling rapid and precise information retrieval during online execution. Our approach employs an LLM-powered offline graph-search algorithm to systematically pre-process low-quality UTGs, converting incomplete or fragmented graphs into structured, vector-based knowledge repositories optimized for retrieval-augmented generation. During online execution, the agent dynamically queries this vector-based repository using embedding-based similarity search, rapidly retrieving relevant navigational paths and app-specific information tailored precisely to the user's intent. This retrieval-augmented approach significantly enhances the agent's decision-making, reducing reliance on extensive real-time exploration and enabling swift, adaptive responses to complex, dynamic app scenarios. Extensive experiments across diverse mobile apps demonstrate that agents equipped with KG-RAG achieve notably higher task completion rates and require fewer steps per task, highlighting KG-RAG's effectiveness in supplementing online agents with critical, domain-specific knowledge.

\noindent The key contributions of our work are:
\begin{itemize} 
\item Proposing \textbf{KG-RAG}, a novel pipeline that transforms incomplete or fragmented UI Transition Graphs (UTGs) into a structured, vector-based knowledge database, optimized for rapid and precise real-time retrieval during online execution. 
\item Introducing KG-Android-Bench and KG-Harmony-Bench, two comprehensive, cross-platform benchmarks tailored for evaluating GUI agents in the diverse Chinese mobile ecosystem.

\item Demonstrating through rigorous evaluation across diverse mobile apps that KG-RAG serves as a plug-and-play module, achieving a 75.8\% task success rate (8.9\% improvement over prior methods) and reducing average task steps from 4.5 to 4.1, significantly enhancing the efficiency and effectiveness of existing GUI agents.
\end{itemize}

\section{Related Work}
Recent LLM-driven agents have made significant progress in automating mobile UI tasks. AutoDroid~\cite{wen2024autodroid}, the most relevant to our work, constructs an app’s UTG offline and uses an LLM online to plan actions, significantly outperforming a GPT-4 baseline in task success rate. However, AutoDroid does not fully leverage the UTG during execution for rapid retrieval of knowledge. 
Other advanced methods, such as Mobile-Agent-v2~\cite{wang2025mobile} and UI-TARS~\cite{qin2025ui}, focus on collaborative agents and end-to-end learning, respectively, achieving strong results on GUI benchmarks. Yet, none of these approaches explicitly leverage a structured knowledge graph of the app’s UI for decision-making. Our proposed KG-RAG fills this gap by providing agents with a UTG-derived knowledge graph, enabling efficient retrieval of navigational knowledge. When integrated with agents like MobileAgent-v2 or UI-TARS, KG-RAG significantly boosts their task success rates, as shown in Section~\ref{subsec:mobileagent}.

Our proposed approach KG-RAG differs from previous GUI agents by integrating structured knowledge graphs derived from UTGs with LLM-based reasoning. While earlier agents either underutilized app-specific graphs or lacked any structured memory, KG-RAG employs a hierarchical knowledge graph, pre-processed for rapid retrieval, to enhance decision-making. This structured memory provides navigation insights that are challenging for LLMs alone to infer. Furthermore, KG-RAG’s plug-and-play design allows it to be seamlessly integrated into various agent architectures, as demonstrated with MobileAgent-v2 and UI-TARS. This flexibility and the combination of offline UTG processing with online retrieval-augmented decision-making mark a significant advancement, leading to higher success rates and greater efficiency in GUI agent tasks.

\section{Method}
This section presents KG-RAG, a framework that enhances online agent decision-making by fully leveraging the rich information embedded within UTGs. An overview of KG-RAG is provided in Figure~\ref{fig:overview}.

\subsection{UTG Extraction}
\label{subsec:utg_extraction}
\begin{figure*}[th]
    \centering
    \includegraphics[width=1\linewidth]{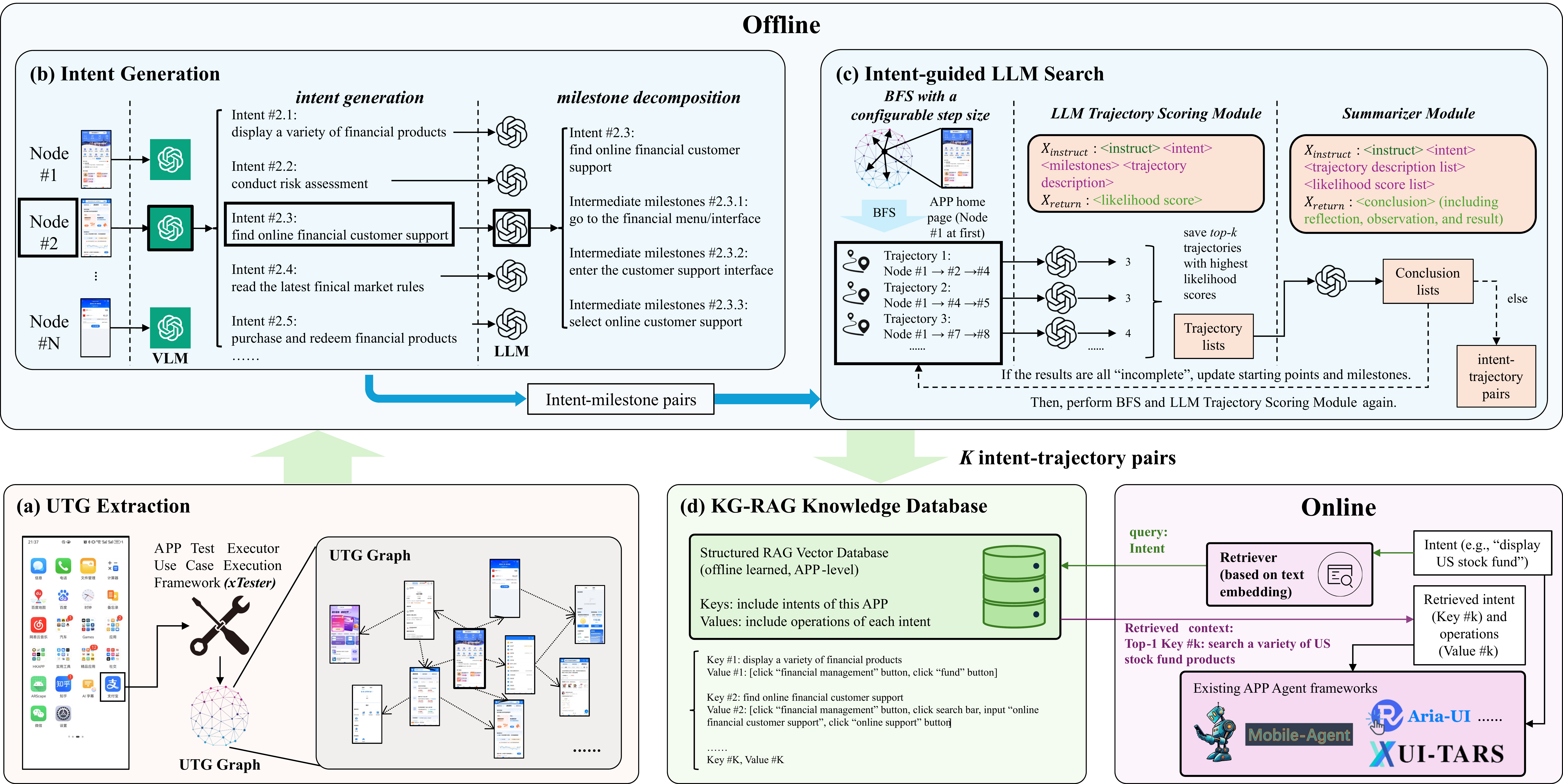}
    \caption{Overview of KG-RAG architecture: \textbf{(a) UTG extraction} capturing app UI navigation structures; \textbf{(b) Intent generation} suggesting plausible user intents and decomposing them into intermediate milestones; \textbf{(c) Intent-guided LLM search} efficiently identifying candidate trajectories aligned with user intents; and \textbf{(d) KG-RAG knowledge database} supporting effective online mobile app interactions.}
    \label{fig:overview}
\end{figure*}

To effectively extract UTGs from mobile applications, we build upon the methodology proposed by DroidBot~\cite{7965248}, adapting it significantly to meet our framework's requirements. Specifically, we introduce a dedicated extraction tool dubbed as \textit{xTester} (App Test Executor Use Case Execution Framework) as shown in Figure~\ref{fig:overview}(a). It is designed to systematically navigate mobile app interfaces, identify interactive UI components, and document their interactions. The outcome is a structured knowledge graph that encompasses the app's UI layouts, control structures, and potential user interactions.

This knowledge graph is represented in a hierarchical JSON format, capturing essential information including app metadata (e.g., product ID, app name, and package name), UI components description, screen description, and actionable interactions associated with specific widgets. Actions, such as swipe gestures, text inputs, and button clicks, are annotated and linked to corresponding UI elements, providing a comprehensive representation of each screen's functionality.

For simpler English-language apps sourced from DroidTask~\cite{wen2024autodroid}, we utilize a 1-hour automated exploration per app using \textit{xTester}. This procedure efficiently captures nearly all relevant app content due to the straightforward UI structures involved. In contrast, for the 30 more complex Chinese-language applications we specifically constructed, we perform an extensive, in-depth exploration for 8 hours. This extended analysis ensures comprehensive coverage of intricate UI states and interactions, significantly surpassing typical extraction methods in both depth and breadth.

Overall, our rigorous UTG extraction approach serves as a crucial foundation for accurately capturing UI widget descriptions and detailed app layout information, directly contributing to the effective construction of the KG-RAG vector database. By providing high-quality and structured vector embeddings of navigational paths, our method enables rapid retrieval and significantly enhanced decision-making capabilities in automated mobile app interactions.

\subsection{Offline Pathfinding via Intent-Guided LLM Search}
The offline pathfinding component is central to KG-RAG and consists of two key parts: \textbf{(1) the intent generation module} and \textbf{(2) the LLM search module}. Together, they enable effective navigation through imperfect UTGs, even with incomplete or missing paths. This component significantly enhances the efficiency and robustness of offline trajectory discovery.

The \textit{intent generation module}, illustrated in Figure~\ref{fig:overview}(b), begins by analyzing screenshots (nodes) extracted from UTGs to identify plausible user intents for each app screen. In practice, we utilize a vision-language model (VLM) to automatically infer these user intents from visual context. Next, a powerful instruction-tuned LLM decomposes each high-level intent into a structured sequence of intermediate milestones. These milestones represent incremental, clearly defined subgoals guiding the agent through complex UI interactions. By breaking down user intents into manageable intermediate steps, our approach reduces ambiguity and prevents premature task termination.

The generated intents and milestones are then utilized by the \textit{LLM trajectory scoring module}, as depicted in Figure~\ref{fig:overview}(c). Given a set of candidate trajectories extracted from the UTG, this module assigns each trajectory both a coarse and a fine-grained score. Specifically, the LLM is used to predict the likelihood of each trajectory successfully achieving the given milestones. The scoring is performed by tokenizing the LLM output, extracting logits corresponding to milestone completion predictions, and computing probability distributions via a softmax operation over the relevant logits indices. 
For a given user intent (with $m$ milestones), we prompt the LLM to predict the likelihood of completing 0 through $m$ milestones along a candidate trajectory. We then apply a softmax over the relevant output logits (Algorithm~\ref{alg:softmax_computation}) to obtain a probability distribution over milestone completion counts. The highest probability in this distribution (corresponding to reaching a certain milestone) is taken as the trajectory’s progress score. 
Next, we compute a proximity score for the trajectory (Algorithm~\ref{alg:proximity_score}) by measuring how closely the LLM’s probability distribution aligns with an ideal monotonically decreasing sequence (favoring trajectories that complete milestones in order). Each trajectory is primarily ranked by its progress score, with the proximity score used to break ties among similar candidates.

Guided by these scoring metrics, we perform a breadth-first search (BFS) to explore the UTG for high-quality trajectories (Algorithm~\ref{alg:llm_bfs_search}). The BFS expands possible paths in increments of the given step size and scores new candidates in batches using the LLM. Any trajectory that fails to achieve a minimum progress (milestone completion) threshold is pruned early, which focuses computation on promising paths. This batched BFS strategy balances thorough exploration with efficiency: it allows parallel LLM evaluations of multiple paths and limits search depth to $max\_depth$. The outcome is a set of top-ranked valid trajectories for each intent. Finally, these trajectories are passed through a summarization module that condenses each sequence of UI actions into a concise description, ready to be stored in the knowledge base for online use.

\begin{algorithm}[t]
\small
\caption{Softmax Computation}
\label{alg:softmax_computation}
\DontPrintSemicolon
\KwIn{
    $x$: Logits array \\
    $start\_ind$: Start index for slicing the logits \\
    $end\_ind$: End index for slicing the logits \\
    $temperature$: Softmax temperature (default $=1$)
}
\KwOut{$softmax\_probs$: Probability distribution over indices $[start\_ind, end\_ind]$}
\BlankLine
\textbf{Procedure:}\\
\Indp
    $x_{\text{slice}} \leftarrow x[\,start\_ind : end\_ind\,]$\;
    $z \leftarrow \exp\!\Big(\frac{x_{\text{slice}}}{\,temperature\,}\Big)$\;
    $softmax\_probs \leftarrow \frac{z}{\sum z}$\;
    \Return $softmax\_probs$\;
\Indm
\end{algorithm}

\begin{algorithm}[ht]
\small 
\caption{Proximity Score Calculation}
\label{alg:proximity_score}
\DontPrintSemicolon
\KwIn{
    $pdf$: Probability distribution function as a list of probabilities
}
\KwOut{
    $proximity\_score$: Scalar value indicating alignment with ideal descending order
}
\BlankLine
\textbf{Procedure:} \\
$ranked\_indices \leftarrow$ indices of $pdf$ sorted in descending order \\
$ideal\_order \leftarrow$ list from $n-1$ down to $0$, where $n$ is length of $pdf$ \\
$proximity\_score \leftarrow -\sum_{i=0}^{n-1} \left( ideal\_order[i] - ranked\_indices[i] \right)^2$ \\
\Return{$proximity\_score$}
\end{algorithm}



\begin{algorithm}[t]
\small
\caption{LLM BFS Search}
\label{alg:llm_bfs_search}
\DontPrintSemicolon
\KwIn{User $intent$, UTG $graph$, $start\_node$, LLM model, $threshold$, $step\_size$, $max\_depth$, top-$K$}
\KwOut{$valid\_trajectories$ achieving the intent}
\textbf{Initialize:} $Q \leftarrow [(start\_node, [start\_node])]$, $valid\_trajectories \leftarrow \emptyset$\;
\While{$Q \neq \emptyset$ \textbf{and} $depth < max\_depth$}{
    $depth \leftarrow depth + step\_size$\;
    Expand $Q$ to generate $candidates$; remove duplicates and loops\;
    \ForEach{batch in $candidates$}{
        Compute LLM scores for batch\;
        \ForEach{trajectory}{
            \If{score $\geq threshold$}{Add to $valid\_trajectories$}
        }
    }
    Keep top-$K$ candidates (by score, prefer shorter paths) and update $Q$\;
}
\Return{$valid\_trajectories$}
\end{algorithm}


\subsection{Knowledge Graph-Augmented Online Execution}
\label{subsec: rag}
During online execution, the KG-RAG agent leverages a vector database of offline-generated intent–trajectory pairs to rapidly retrieve relevant navigation knowledge. 
Each entry in our structured vector database is a key–value pair: the key is a high-dimensional embedding of an intent discovered offline, and the value is the corresponding trajectory (sequence of UI actions to fulfill that intent). These intent–trajectory pairs are obtained from the offline phase: the VLM infers plausible user intents from app screens (Figure~\ref{fig:overview}(b)), and the LLM-based search finds successful trajectories for those intents (Figure~\ref{fig:overview}(c)). We encode each intent (along with its trajectory) into a vector using a text embedding model, creating a repository of navigational knowledge optimized for fast similarity search.

During online execution, the agent leverages this repository to retrieve relevant guidance in real time. The user’s current task instruction is encoded into a query vector (using the same text embedding model) and compared against the stored intent vectors via cosine similarity. The agent then retrieves the top-$K$ most similar intent–trajectory entries (Figure~\ref{fig:overview}(d)). Each retrieved trajectory serves as contextual knowledge, suggesting a proven navigation path for the agent to follow. For example, as illustrated in Figure~\ref{fig:overview}(d), if the user needs to “display US stock fund” information, the agent will retrieve the stored trajectory that accomplishes this task, providing the sequence of UI actions required to reach the stock fund content. By following such retrieved trajectories, the KG-RAG-enhanced agent can quickly navigate to the target state instead of relying on trial-and-error exploration. This retrieval-augmented execution allows the agent to respond adaptively to new tasks using the collective knowledge encoded in the UTG-derived graph memory.

\begin{table}[t]
    \caption{Comparison of KG-Android-Bench and KG-Harmony-Bench with existing GUI agent benchmarks}
    \centering
    \setlength{\tabcolsep}{0.5pt}
    \small
    \begin{tabular}{l|c|c}
        \hline
        \textbf{Benchmark} & \textbf{No. of Tasks} & \textbf{No. of Apps} \\
        \hline
        AndroidLab & 138 & 9 \\
        DroidTask & 162 & 12  \\
        \textbf{KG-Android-Bench (Ours)} & 300 & 30  \\
        \textbf{KG-Harmony-Bench (Ours)} & 150 & 15  \\
        \hline
    \end{tabular}
    \label{tab:benchmark_comparison}
\end{table}

\begin{table}[t]
\caption{Applications in KG-Android-Bench covering 10 functional categories.}
\centering
\small
\resizebox{0.5\textwidth}{!}{ 
\begin{tabular}{ll}
\toprule
\textbf{Category} & \textbf{Applications} \\
\midrule
Music \& Audio & QQ Music, NetEase Cloud Music, Himalaya FM \\
Video \& Entertainment & Douyin(Chinese Tiktok), Youku, Douyu, WeSing \\
Social \& Communication & Weibo, Zhihu, Baihe \\
Navigation \& Travel & Amap, Ctrip, Zhixing Train Tickets \\
E-commerce \& Retail & Taobao, Vipshop, Dianping \\
Food Services & Meituan Takeaway, Pupu Supermarket \\
Health \& Lifestyle & Keep, Moji Weather, Daily Alarm Clock \\
News \& Reading & Tomato Novel, Jinri Toutiao,  Hupu, Dongchedi \\
Productivity & Baidu Browser, NetEase Mail, Youdao Dictionary \\
Photo \& Video Editing & Xingtu, CapCut \\
\bottomrule
\end{tabular}
}
\label{tab:app_list}
\end{table}

\begin{table*}[th]
\caption{Per-application comparison of AutoDroid vs.\ KG-RAG using GPT-4 model. SR and DA are in \%. AS is the average number of steps taken to complete a task.}
\centering
\small
\resizebox{0.95\textwidth}{!}{%
\begin{tabular}{lcccccc}
\toprule
\textbf{App} & \textbf{AutoDroid SR}$\uparrow$ & \textbf{KG-RAG SR}$\uparrow$ & \textbf{AutoDroid DA}$\uparrow$ & \textbf{KG-RAG DA}$\uparrow$ & \textbf{AutoDroid AS}$\downarrow$ & \textbf{KG-RAG AS}$\downarrow$ \\
\midrule
Gallery        & 40.00 & 60.00 & 40.00 & 60.00 & 3.75 & 3.25 \\
Music Player   & 80.00 & 80.00 & 80.00 & 90.00 & 3.50 & 3.25 \\
Voice Recorder & 80.00 & 90.00 & 80.00 & 100.00 & 4.13 & 3.88 \\
Dialer         & 80.00 & 86.67 & 93.33 & 100.00 & 5.50 & 4.92 \\
Contacts       & 73.33 & 93.33    & 80.00    & 93.33     & 4.64 & 4.45 \\
Calendar       & 50.00 & 56.25 & 75.00 & 81.25  & 4.25 & 4.00 \\
Notes          & 60.00 & 73.33 & 73.33 & 80.00  & 5.44 & 5.11 \\
SMS Messenger  & 80.00 & 80.00 & 86.67 & 86.67  & 4.45 & 4.18 \\
File Manager   & 60.00 & 66.67 & 80.00 & 73.33  & 3.44 & 3.22 \\
Clock          & 73.33 & 80.00 & 80.00 & 93.33  & 4.18 & 3.36 \\
App Launcher   & 80.00 & 90.00 & 90.00 & 90.00  & 7.00 & 6.25 \\
Camera         & 46.67 & 53.33 & 60.00 & 66.67  & 3.57 & 3.29 \\
\midrule
\textbf{Average} & 66.94 & \textbf{75.80} & 76.53 & \textbf{84.55} & 4.49 & \textbf{4.10} \\
\bottomrule
\end{tabular}}
\label{tab:autodroid_detailed_comparison}
\end{table*}

\begin{table*}[th]
\caption{Performance comparison of AutoDroid vs.\ KG-RAG using Qwen2-VL-72B model.}
\centering
\small
\resizebox{0.95\textwidth}{!}{%
\begin{tabular}{lcccccc}
\toprule
\textbf{App} & \textbf{AutoDroid SR}$\uparrow$ & \textbf{KG-RAG SR}$\uparrow$ & \textbf{AutoDroid DA}$\uparrow$ & \textbf{KG-RAG DA}$\uparrow$ & \textbf{AutoDroid AS}$\downarrow$ & \textbf{KG-RAG AS}$\downarrow$ \\
\midrule
\textbf{Average} & 62.8 & \textbf{70.5} & 71.6 & \textbf{80.2} & 8.7 & \textbf{7.9} \\
\bottomrule
\end{tabular}}
\label{tab:autodroid_comparison}
\end{table*}

\section{Benchmark Construction}
We present \textbf{KG-Android-Bench} and \textbf{KG-Harmony-Bench}, two comprehensive cross-platform benchmarks for evaluating GUI agents in Chinese mobile ecosystems. As shown in Table~\ref{tab:benchmark_comparison}, KG-Android-Bench significantly outperforms existing benchmarks with 300 tasks across 30 mainstream Chinese applications, compared to DroidTask’s 162 tasks and AndroidLab’s 138 tasks. This expanded scale, coupled with support for 10 functional categories (as shown in Table~\ref{tab:app_list}), offers a more diverse and thorough evaluation of agent capabilities across varied mobile environments.

A key innovation of \textbf{KG-Android-Bench} is its use of structured knowledge graphs and intent-action mappings, which provide a detailed map of app interfaces. These elements enable more realistic assessments of agents' performance on high-frequency mobile interactions, such as those found in social media, e-commerce, navigation, and fitness tracking. Unlike existing benchmarks that rely solely on basic UI traces, KG-Android-Bench encodes the full navigation flow and task execution sequences within each app.

For instance, to complete the task ``View App Privacy Policy'' in the Tomato Novel app shown in Figure~\ref{fig:motivation}, KG-Android-Bench defines a sequence of the following actions: \textbf{Step 1: Open Profile} by tapping the \textbf{``My Profile''} button (the user’s profile/account section). \textbf{Step 2: Open Settings} by tapping the \textbf{``Settings'' }gear icon. \textbf{Step 3: About Section} by tapping \textbf{``About Tomato''} (the about page of the app). \textbf{Step 4: Privacy Policy} by tapping \textbf{``Privacy Policy''} to view the content of the policy.

These steps are generated by KG-RAG, as demonstrated in Figure~\ref{fig:motivation}, where the agent successfully identifies the correct navigation path. Such sequences capture multi-step relationships and guides the agent’s decision-making process in real-time. By leveraging this structured data, agents can efficiently navigate deeply nested app structures and complete multi-step tasks without the need of extensive random exploration.

A knowledge graph enhances a GUI agent’s ability to navigate deep UI structures by capturing multi-step relationships. In KG-Android-Bench, each app’s actions are linked in a logical sequence, allowing the agent to retrieve the correct next step for a given intent. This knowledge-driven approach enables effective multi-step reasoning for complex interactions that simpler benchmarks might fail to capture.

By leveraging the knowledge graph as context, a GUI agent can navigate deep or hidden UI elements through graph-based retrieval. The combination of detailed intent-action mapping and structured knowledge allows the agent to anticipate how to achieve high-level goals, such as locating a privacy policy or completing a purchase, by following a sequence of UI actions. This significantly improves the agent’s ability to handle multi-step tasks and adapt to app-specific workflows.

Furthermore, KG-Android-Bench and KG-Harmony-Bench are designed to support both Android and HarmonyOS, respectively, ensuring consistent evaluation across different mobile operating systems. This cross-platform capability is especially important given the rising adoption of HarmonyOS in China, now accounting for approximately 17\% of the domestic mobile OS market. By evaluating agents on the same tasks in both environments, KG-RAG ensures robust generalization across platform-specific UI differences.

In summary,  we set new benchmarks called KG-Android-Bench and KG-Harmony-Bench for evaluating autonomous mobile agents by integrating structured knowledge graphs and detailed intent-action mappings. Its cross-platform support makes it an essential tool for assessing agents in diverse, real-world app environments, driving further advancements in the field.
\textcolor{black}{Compared with prior Android datasets that mainly provide UI traces, our KG-Android-Bench and KG-Harmony-Bench couple \emph{intent--action} mappings with UTG-derived graph memory, enabling evaluation of multi-step, deeply nested goals common in Chinese mobile apps (e.g., finance, e-commerce, travel). The paired Android/HarmonyOS suites also stress \emph{cross-platform robustness} under heterogeneous UI paradigms---a setting underrepresented in existing resources---and therefore serve as a testbed for graph-augmented agents rather than a near-duplicate of earlier benchmarks.}

\begin{figure*}[t]
\centering
\includegraphics[width=0.96\linewidth]{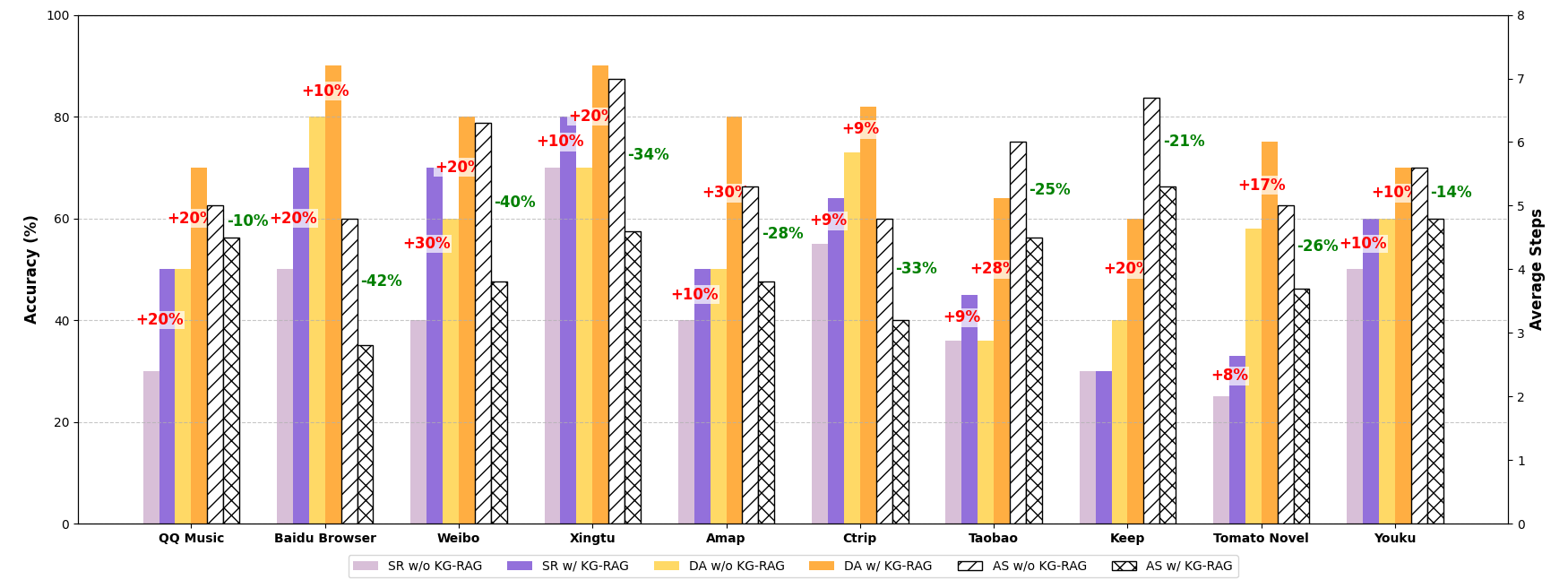}
\caption{Performance comparison across 10 representative Chinese mobile applications, showing improvements in Success Rate (SR) and Decision Accuracy (DA) with KG-RAG, while reducing task average steps (AS) (indicated by striped bars). Red numbers reflect accuracy gains, and green numbers indicate step reductions.}
\label{fig:kg_rag_comparison}
\end{figure*}



\begin{table*}[t]
\caption{Performance comparison of MobileAgent-v2 and UI-TARS with and without the integration of KG-RAG on the ``QQ Music'' application.}
\centering
\small
\setlength{\tabcolsep}{4.5pt}
\begin{tabular}{llcccc}
\toprule
\textbf{Method} & \textbf{Perception Model} & \textbf{Decision Model} & \textbf{SR (\%)}$\uparrow$  & \textbf{DA (\%)}$\uparrow$ & \textbf{AS}$\downarrow$  \\
\midrule
MobileAgent-v2 & GroundingDINO & Qwen2-VL & 20.0 & 40.0 & 5.5 \\
MobileAgent-v2+\textbf{KG-RAG} & GroundingDINO & Qwen2-VL & \textbf{30.0} & \textbf{60.0}  & \textbf{5.0} \\
\midrule
MobileAgent-v2 & VUT & Qwen2-VL & 30.0 & 50.0 & 5.0 \\
MobileAgent-v2+\textbf{KG-RAG} & VUT & Qwen2-VL & \textbf{50.0} & \textbf{70.0}  & \textbf{4.5} \\
\midrule
MobileAgent-v2 & GroundingDINO & GPT-4o & 30.0 & 50.0 & 5.3 \\
MobileAgent-v2+\textbf{KG-RAG} & GroundingDINO & GPT-4o & \textbf{40.0} & \textbf{60.0} & \textbf{5.0} \\
\midrule
MobileAgent-v2 & VUT & GPT-4o & 50.0 & 60.0 & 4.8 \\
MobileAgent-v2+\textbf{KG-RAG} & VUT & GPT-4o & \textbf{60.0} & \textbf{70.0} & \textbf{3.2} \\
\midrule
UI-TARS & UI-TARS-7B-SFT & UI-TARS-7B-SFT & 90.0 & 90.0 & 5.9 \\
UI-TARS+\textbf{KG-RAG} & UI-TARS-7B-SFT & UI-TARS-7B-SFT & 90.0 & \textbf{100.0} & \textbf{5.2} \\
\bottomrule
\end{tabular}
\label{tab:mobileagentv2_comparison}
\vspace{-4mm}
\end{table*}

\section{Experiments}
\label{sec:experiments}

\subsection{Experimental Setup}
\label{subsec:setup}
\paragraph{Evaluation Benchmarks.} We conduct comprehensive evaluations on the following three benchmarks:
(1) \textbf{DroidTask (EN)~\cite{wen2024autodroid}}: 162 tasks across 12 English Android apps (e.g., Gallery, Camera, and File Manager.)
(2) \textbf{KG-Android-Bench (CN)} (Ours):  A new benchmark of over 200 real-world tasks spanning 30 Chinese apps. These tasks cover diverse domains such as music, social media, navigation, e-commerce, and more (in Chinese UI environments).
(3) \textbf{KG-Harmony-Bench (CN)} (Ours): A set of 150 cross-platform tasks on HarmonyOS devices, covering 15 app categories (e.g., mapping apps like Amap, travel apps like Ctrip). This benchmark evaluates an agent’s ability to generalize across platforms.

To evaluate agent performance, we use three key metrics: Success Rate (SR), which measures the percentage of user instructions successfully completed; Decision Accuracy (DA), reflecting the correctness of the agent’s action decisions; and Average Steps (AS), indicating task efficiency through the average number of steps taken to finish a task.

\subsection{Comparison with AutoDroid}
\label{subsec:autodroid} 
We first compare our KG-RAG approach with similar UTG-based AutoDroid~\cite{wen2024autodroid} on the DroidTask benchmark using two LLM backends: Qwen2-VL~\cite{wang2024qwen2} and GPT-4~\cite{achiam2023gpt}. 

First, Table~\ref{tab:autodroid_detailed_comparison} provides a detailed per-application performance breakdown for AutoDroid and KG-RAG on the DroidTask benchmark (using a GPT-4 backend). We report Success Rate (SR), Decision Accuracy (DA), and Average Steps (AS) for each app. KG-RAG achieves consistent improvements on almost all individual apps, underlining the robustness of our approach. For example, on the Gallery app, KG-RAG raises the success rate from 40.0\% to 60.0\% and the decision accuracy from 40.0\% to 60.0\%, while also reducing the average steps from 3.75 to 3.25. Similar gains are observed in most cases (e.g., Voice Recorder SR 80\%$\rightarrow$90\%, Dialer DA 93.3\%$\rightarrow$100\%), indicating that our method’s advantages hold at the per-app level, not just in aggregate results.

Additionally, Tables \ref{tab:autodroid_detailed_comparison} and \ref{tab:autodroid_comparison} summarize the overall performance of AutoDroid vs. KG-RAG across key metrics and across two different LLM backends (GPT-4 and Qwen2-VL). KG-RAG consistently outperforms AutoDroid on every metric under both LLM settings. In particular, KG-RAG yields higher average success rates and decision accuracy, while requiring fewer steps on average. These results demonstrate that our knowledge-augmented approach provides clear benefits over the prior UTG-based method in both accuracy and efficiency.

\subsection{Evaluating KG-RAG as a Plug-and-Play Module in GUI Agents}
\label{subsec:mobileagent}
To demonstrate the versatility of our approach, we integrate KG-RAG as a plug-and-play module into two state-of-the-art GUI agent frameworks: MobileAgent-v2~\cite{wang2025mobile} and UI-TARS~\cite{qin2025ui} and we choose different perception models (VUT~\cite{li2021vut} and GroundingDINO~\cite{liu2024grounding}) and decision models (GPT4-o~\cite{hurst2024gpt} and Qwen2-VL~\cite{wang2024qwen2}). This integration is seamless, requiring no architectural changes, highlighting that KG-RAG can serve as a general enhancement layer for different agents.

Table~\ref{tab:mobileagentv2_comparison} summarizes the performance of MobileAgent-v2 and UI-TARS with and without KG-RAG. For MobileAgent-v2, KG-RAG boosts the success rate (SR) and the decision accuracy (DA), while also reducing the number of steps needed for task completion. Similarly, UI-TARS benefits from the KG-RAG integration, showing notable improvements in both SR and DA.

Figure~\ref{fig:kg_rag_comparison} illustrates these improvements, showing that agents equipped with KG-RAG not only achieve higher success and accuracy, but also require fewer interactions to complete tasks compared to their base versions. The consistent performance boost across both MobileAgent-v2 and UI-TARS emphasizes KG-RAG’s effectiveness as a drop-in module. By providing relevant contextual knowledge on the fly, KG-RAG helps guide the agent’s decisions, leading to more successful outcomes and streamlined action sequences.

\begin{table}[t]
\centering
\small
\caption{Performance comparison of UI-TARS and MobileAgent-v2 using GPT-4o on HarmonyOS.}
\resizebox{0.48\textwidth}{!}{ 
\setlength{\tabcolsep}{4.5pt}
\begin{tabular}{lccc}
\toprule
\textbf{Method} & \textbf{SR (\%)}$\uparrow$ & \textbf{DA (\%)}$\uparrow$  & \textbf{AS}$\downarrow$ \\
\midrule
MobileAgent-v2 & 41.9 & 64.2  & 4.6 \\
MobileAgent-v2 + KG-RAG & \textbf{55.4} & \textbf{77.4} & \textbf{4.2} \\
\midrule
UI-TARS & 61.4 & 71.0  & 4.4 \\
UI-TARS + KG-RAG & \textbf{70.9} & \textbf{85.5}  & \textbf{4.0} \\
\bottomrule
\end{tabular}
}
\label{tab:harmony_comparison}
\end{table}

\subsection{Evaluation on HarmonyOS}
\label{subsec:harmonyos}
Table~\ref{tab:harmony_comparison} reports similar success rates and decision accuracy for KG-RAG on HarmonyOS as the results on Android in Table~\ref{tab:mobileagentv2_comparison}, showing consistent performance across both platforms. This highlights KG-RAG's ability to generalize across different UI paradigms without requiring platform-specific modifications, making it a versatile solution for mobile app interactions.

\subsection{Ablation Study of RAG Construction}
To evaluate the impact of different knowledge construction components, we conduct an ablation study varying the choice of VLM used in intent generation module in Figure~\ref{fig:overview}(b) and LLM used in the LLM search module in Figure~\ref{fig:overview}(c) to construct the KG-RAG Knowledge Database.

Table~\ref{tab:rag_ablation_android} compares four configurations of VLM and LLM used to construct the KG-RAG knowledge database on KG-Android-Bench (using UI-TARS as the agent).
For the VLM+LLM configurations, we experiment with InternVL2~\cite{chen2024internvl}, Qwen2-VL~\cite{wang2024qwen2vl}, and DeepSeek-14B~\cite{deepseekai2025deepseekr1incentivizingreasoningcapability}.
The results show that leveraging a stronger vision-language model (VLM) and a specialized reasoning LLM yields the best performance. In particular, Qwen2-VL-72B + DeepSeek-14B achieves the highest success rate (SR 78.33\%) and decision accuracy (DA 82.03\%), along with the lowest average steps (AS 4.92), outperforming all other combinations. 
Overall, the Qwen2-VL + DeepSeek-14B pairing produces the largest gains in both success and accuracy, and also finds shorter navigation paths (lowest AS), confirming that richer visual semantics combined with focused reasoning yields the most effective knowledge graph construction for KG-RAG.

We further investigate different text embedding models for retrieval, using doubao-embedding-text~\cite{seed2025seed}, gte-multilingual-base~\cite{zhang2024mgte}, and multilingual-e5-large-instruct~\cite{wang2024multilingual}. Table~\ref{tab:embedding_ablation} examines the impact of different text embedding models on retrieval performance. Here, doubao-embedding-text~\cite{seed2025seed} emerges as the top performer, achieving the highest SR (73.90\%) and a joint-highest DA (80.00\%), while also requiring the fewest steps (AS 4.73). Based on these results, we select doubao-embedding-text for the final KG-RAG system due to its stronger retrieval alignment and lower step count, which together contribute to higher overall efficiency.


\begin{table}[t]
\caption{Effect of VLM and LLM choices on KG-RAG database construction performance on KG-Android-Bench.}
\resizebox{0.48\textwidth}{!}{ 
\centering\small
\setlength{\tabcolsep}{4.0pt}
\begin{tabular}{lccc}
\toprule
\textbf{VLM + LLM Combination}& \textbf{SR (\%)}$\uparrow$ & \textbf{DA (\%)}$\uparrow$  & \textbf{AS}$\downarrow$ \\
\midrule
InternVL2-76B+Qwen2-72B & 67.00 & 70.70 & 5.00 \\ %
InternVL2-76B+Deepseek-14B & 70.96 & 74.66 & 5.01 \\
Qwen2-VL-72B+Qwen2-72B & 72.59 & 78.07 & 5.16 \\
Qwen2-VL-72B+Deepseek-14B & \textbf{78.33} & \textbf{82.03} & \textbf{4.92} \\
\bottomrule
\end{tabular}
}
\label{tab:rag_ablation_android}
\end{table}

\begin{table}[t]
\caption{Comparison of retrieval performance using different text embedding models on KG-Android-Bench.}
\resizebox{0.48\textwidth}{!}{ 
\centering\small
\setlength{\tabcolsep}{4.0pt}
\begin{tabular}{lccc}
\toprule
\textbf{Embedding Model}& \textbf{SR (\%)}$\uparrow$ & \textbf{DA (\%)}$\uparrow$  & \textbf{AS}$\downarrow$ \\
\midrule
multilingual-e5-large-instruct & 70.60 & 76.60 & 5.08 \\ 
gte-multilingual-base & 70.60 & 80.00 & 5.19 \\ 
doubao-embedding-text & \textbf{73.90} & \textbf{80.00} & \textbf{4.73} \\ 
\bottomrule
\end{tabular}
}
\label{tab:embedding_ablation}
\end{table}

\subsection{Generalization Across GUIs and UTG Cost Trade-offs}
\label{sec:gen-cost}
\textcolor{black}{We further report results that (1) transfer KG-RAG to non-mobile GUIs without retraining, (2) test cross-device/OS robustness, and (3) quantify the UTG construction cost–quality trade-off.
First, we transfer KG-RAG to \emph{non-mobile} GUIs without any retraining; 
Table~\ref{tab:web-desktop} shows large gains on Weibo (web) and QQ Music (desktop). 
Second, we evaluate \emph{cross-device/OS} robustness on \emph{Weibo}; 
Table~\ref{tab:cross-device} demonstrates consistent improvements from low-end to flagship devices as well as on HarmonyOS. 
Finally, we quantify the \emph{UTG construction cost--quality} trade-off; 
Table~\ref{tab:utg-ablation} indicates accuracy saturates at about 4 hours per complex app, enabling practical deployment with bounded offline cost.}

As seen in Table~\ref{tab:web-desktop}, transferring a mobile-built KG\textendash RAG to web/desktop yields +40\% SR on Weibo-web and +20\% SR on QQ Music-desktop, without any platform-specific retraining. 
Table~\ref{tab:cross-device} further shows that the gains hold across heterogeneous chipsets and on HarmonyOS, while steps (AS) are consistently reduced. 
Finally, Table~\ref{tab:utg-ablation} reveals that performance saturates at about 4 hours of UTG extraction, indicating a practical operating point for large-scale deployments.

\begin{table}[t]
\centering
\caption{Non-mobile GUIs via training-free transfer of a mobile-built KG-RAG database.}
\label{tab:web-desktop}
\resizebox{\columnwidth}{!}{
\begin{tabular}{l l c c c}
\toprule
Domain & Method & SR(\%)$\uparrow$ & DA(\%)$\uparrow$ & AS$\downarrow$ \\
\midrule
Weibo (web)         & UI-TARS-web            & 50.0 & 70.0  & 7.6 \\
                    & \quad + KG-RAG         & \textbf{90.0} & \textbf{100.0} & \textbf{5.2} \\
\hline
QQ Music (desktop)  & UI-TARS-desktop        & 60.0 & 60.0  & 5.9 \\
                    & \quad + KG-RAG         & \textbf{80.0} & \textbf{90.0}  & \textbf{4.3} \\
\bottomrule
\end{tabular}
}
\end{table}

\begin{table}[t]
\centering
\caption{Cross-device/OS performance on \emph{Weibo}.}
\label{tab:cross-device}
\begingroup
\setlength{\tabcolsep}{3.2pt}
\renewcommand{\arraystretch}{0.90}
\resizebox{\columnwidth}{!}{
\begin{tabular}{l l c c c}
\toprule
OS & Device (Chip) & SR(\%)$\uparrow$ & DA(\%)$\uparrow$ & AS$\downarrow$ \\
\midrule
Baseline (no KG-RAG) & HUAWEI P40 (Kirin 990)        & 40.0 & 60.0 & 6.25 \\
\hline
Android               & HUAWEI Y9s (Kirin 710F)       & 60.0 & 70.0 & 5.75 \\
Android               & OPPO K9s (Snapdragon 778G)    & 60.0 & 80.0 & 5.00 \\
Android               & Vivo iQOO 8 (Snapdragon 888)  & 70.0 & 70.0 & 4.50 \\
Android               & HUAWEI P40 (Kirin 990)        & 70.0 & 80.0 & 3.75 \\
HarmonyOS             & HUAWEI Mate 60 (Kirin 9000S)  & 80.0 & 80.0 & 3.25 \\
\bottomrule
\end{tabular}
}
\endgroup
\end{table}

\begin{table}[t]
\centering
\caption{UTG extraction time vs.\ performance on \emph{Weibo}. Accuracy saturates at $\sim$4h.}
\label{tab:utg-ablation}
\resizebox{\columnwidth}{!}{
\begin{tabular}{l c c c c}
\toprule
Extraction Time & SR(\%)$\uparrow$ & DA(\%)$\uparrow$ & AS$\downarrow$ & $\Delta$SR over Baseline \\
\midrule
Baseline (w/o RAG) & 40.0 & 60.0 & 6.25 & -- \\
\hline
1 h                & 50.0 & 60.0 & 5.75 & +25\% \\
2 h                & 60.0 & 70.0 & 5.25 & +50\% \\
4 h                & 70.0 & 80.0 & 4.50 & +75\% \\
8 h                & 70.0 & 80.0 & 3.75 & +75\% \\
\bottomrule
\end{tabular}
}
\vspace{-1.2mm}
\end{table}

\section{Conclusion}
The paper presents KG-RAG, a framework that enhances GUI agents by leveraging structured knowledge from UTGs. KG-RAG addresses key limitations of existing agents, such as their inability to fully utilize app-specific knowledge from incomplete UTGs. Our intent-guided offline pathfinding algorithm transforms UTGs into structured vector embeddings, creating a robust knowledge database that improves real-time decision-making with pre-computed navigation paths for specific user intents.

We also introduce KG-Android-Bench, a comprehensive benchmark for GUI agents, supporting cross-platform (Android, HarmonyOS) with detailed intent-action mappings in knowledge graphs. Experimental results show that KG-RAG significantly improves agent efficiency, reducing task completion steps and increasing success rates. Our findings highlight the effectiveness of knowledge-driven retrieval augmentation in overcoming practical challenges for mobile GUI agents, setting new performance benchmarks.

\section*{Limitations}
KG-RAG relies on app UI Transition Graphs (UTG) extracted by an automatic app testing system. Currently, it costs one or several hours to fully construct the UTG due to the ambitious goal of obtaining a high page coverage inside the app. In the future, it is worth exploring a more advanced automatic app testing system to mitigate resource cost. Moreover, it is interesting to adopt this KG-RAG framework in the domain of web and PC.

Furthermore, KG-RAG designs a knowledge database for each app. Future work could explore the design of a vertical domain (e.g., shopping) .



\section*{Ethics Discussion}

In constructing the knowledge graph and developing the KG-RAG system, we took careful steps to protect user privacy and data confidentiality:
(1) \textbf{Anonymous Data Collection:}  All UTGs are fully anonymized and contain only abstract UI structures and navigation paths, without any PII or user content;
(2) \textbf{Automatic Privacy Masking:} Potentially sensitive on-screen fields are automatically masked during data capture;
(3) \textbf{No Authentication Required:} We evaluate only logged-out scenarios and never use accounts, passwords, or authenticated actions;
(4) \textbf{Data Security:} Data are stored securely for research purposes only, and no information that can identify individuals is released.

\clearpage
\bibliography{custom}




\end{document}